\documentclass[12pt]{elsarticle}

\usepackage{graphicx}

\usepackage{amssymb}

\newcommand{\etal}{{\it et\ al}.}

\begin{document}

\begin{frontmatter}





\title{High-Speed Charge-to-Time Converter ASIC for the
 Super-Kamiokande Detector}

\author[icrr]{H.~Nishino\corref{cor}\fnref{fn1}}
\ead{nishino@post.kek.jp}
\author[icrr]{K.~Awai}
\author[icrr]{Y.~Hayato}
\author[icrr]{S.~Nakayama}
\author[icrr]{K.~Okumura}
\author[icrr]{M.~Shiozawa}
\author[icrr]{A.~Takeda}
\author[iwatsu]{K.~Ishikawa}
\author[iwatsu]{A.~Minegishi}
\author[kek]{Y.~Arai}

\cortext[cor]{Corresponding author. Tel.: +81 471 36 3130; fax: +81 471 36 3126.}
\address[icrr]{Institute for Cosmic Ray Research, University of Tokyo,
 Chiba 277-8582, Japan}
\address[iwatsu]{Iwatsu Test Instruments Corporation, Tokyo 168-8511, Japan}
\address[kek]{The Institute of Particle and Nuclear Studies, KEK,
 Ibaraki 305-0801, Japan}
\fntext[fn1]{Present address: The Institute of Particle and Nuclear Studies, KEK,
 Ibaraki 305-0801, Japan}

\begin{abstract}
A new application-specific integrated circuit (ASIC),
the high-speed charge-to-time converter
(QTC) IWATSU CLC101, provides three channels,
each consisting of
preamplifier, discriminator, low-pass filter, and charge integration
circuitry,
optimized for the waveform of a photomultiplier tube (PMT).
This ASIC detects PMT signals using individual built-in discriminators
and drives output timing signals whose width represents
the integrated charge of the PMT signal.
Combined with external input circuits composed of passive elements,
the QTC provides full analog signal processing 
for the detector's PMTs, ready for further processing by
time-to-digital converters (TDCs).
High-rate ($>$1~MHz) signal processing is achieved by
short-charge-conversion-time
and baseline-restoration circuits.
Wide-range charge measurements are enabled by offering
three gain ranges while maintaining a short cycle time.
QTC chip test results show good analog performance, 
with efficient detection for a single photoelectron signal, 
4 orders of magnitude dynamic range (0.3~mV $\sim$ 3~V;
0.2 $\sim$ 2500~pC), 1\% charge linearity, 0.2~pC charge resolution,
and 0.1~ns timing resolution.
Test results on ambient temperature dependence, channel
isolation, and rate dependence also meet specifications.
\end{abstract}

\begin{keyword}
Super-Kamiokande \sep
photomultiplier tube readout \sep
charge-to-time converter \sep
self-trigger \sep
high-speed \sep
low-noise

\end{keyword}
\end{frontmatter}


 \section{Introduction}
 Super-Kamiokande (SK) is the world's largest ring-imaging water
Cherenkov detector for astroparticle and elementary particle physics.
In operation since 1996, it has achieved several important
scientific results, notably discovery of neutrino flavor mixings and
their masses~\cite{PRL_81_1562, PRD_69_011104}.  The detector will
stay in operation to further
explore atmospheric and solar neutrinos, 
artificial neutrinos in an accelerator-based long baseline neutrino
experiment~\cite{hep-ex_0106019}, cosmic neutrinos from supernova
explosions~\cite{PRL_90_061101}, 
baryon-number-violated proton decay signals~\cite{PRL_81_3319},
and more.

\begin{figure}
 \centering
 \includegraphics[width=13cm]{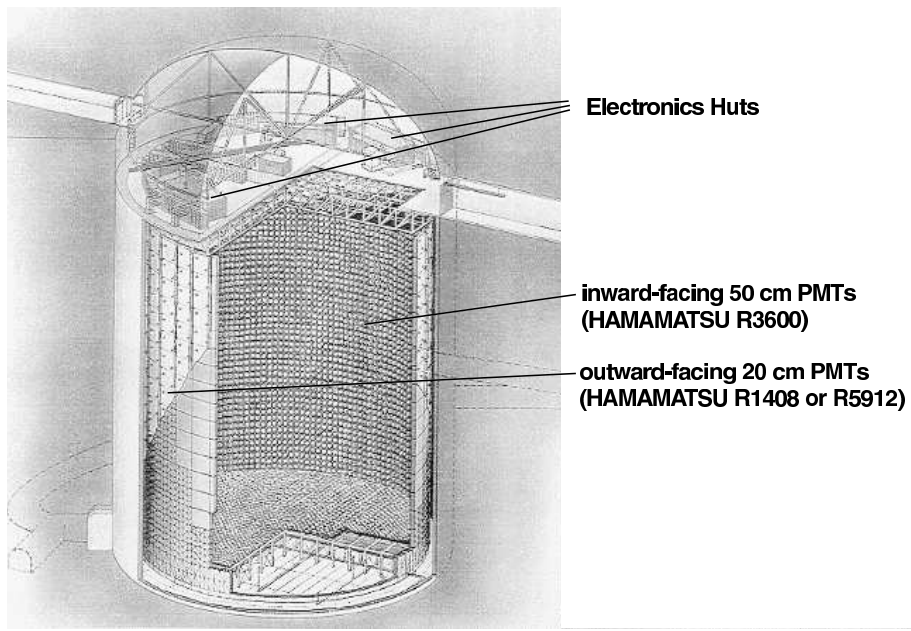}
 \caption{Sketch of the Super-Kamiokande detector.}
 \label{SK_detector}
\end{figure}

The 50 kt cylindrical water tank shown in Fig.~\ref{SK_detector}, 
39 m in diameter and
42 m in height, houses an inward-facing array of 50-cm
photomultiplier tubes (PMTs) and outward-facing array of 20-cm anticounter
PMTs.  The former array consists of 11,129 HAMAMATSU R3600
hemispherical PMTs; 
the latter array consists of 611 R1408 and 1274 R5912 PMTs. 
 Details of the detector and its data acquisition electronics
are reported elsewhere~\cite{NIMA_501_418, NIMA_320_310}.
The SK collaboration decided to replace detector's front-end electronics
with new ones,
which motivated the development of a new device as described herein.

It is required that front-end electronics for the detector record the 
integrated
charge of the PMT signals and their timing, corresponding to the amount of
detected Cherenkov light and its time of arrival on the PMT surface,
respectively.  Each channel self-triggers$-$that is,
starts charge integration by itself using its own
discriminator$-$because an external start timing signal is not available
for natural neutrino and proton decay observations.  The electronics 
also match the characteristics of the PMTs; the R3600
PMTs operate with a gain of 10$^7$ and their signals are transported to
the electronics via 70-m RG58 coaxial cables.  
The charge dynamic range of the PMT is $\sim$ 1000~photoelectron 
(p.e.; plural p.e.s). 

\begin{figure}
 \centering
 \includegraphics[width=6.5cm]{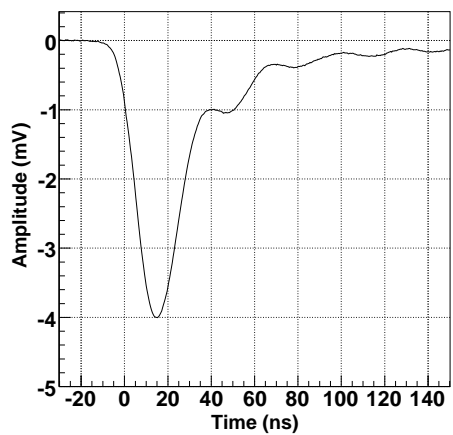}
 \caption{
 Waveform of single photoelectron signals of the R3600 PMT
 recorded at the end of a 70-m RG58 coaxial cable.
 }
 \label{pmtwaveform_1pe}
\end{figure}

Figure \ref{pmtwaveform_1pe} shows the waveform of a single p.e.  
signal from an R3600 PMT.
Charge and timing resolution of the PMT at the single p.e. level
($\simeq2$~pC) are about 100\% and 2 $-$ 3~ns,
respectively~\cite{NIMA_329_299},
and the single p.e. signal should be detected with high efficiency at
a discriminator threshold of $\sim$ 1/4~p.e.s.  
A short ($<$ 1~$\mu$sec) cycle time is 
required to record, with high efficiency, two successive events 
in which Michel electron events follow muon events with a mean interval of
the muon's lifetime, 2~$\mu$sec.

For long-term operation of the SK detector, an integrated
3-channel device for PMT signal readout has been developed.  This
application-specific integrated circuit (ASIC), called the high-speed 
charge-to-time converter (QTC) IWATSU CLC101, was designed 
in CMOS 0.35~$\mu$m technology.  
It has built-in discriminators to trigger its integration
circuits by itself, and encodes the amount of input charge to 
the timing signal,
where leading edge and width represent timing and integrated charge of
the input signal, respectively.  Necessary preamplifier and analog delay
circuits also reside in the ASIC to enable high-density
application.  

The new ASIC offers three charge dynamic ranges,
each with a short cycle time ($<$ 1~$\mu$sec).  
In real application, the output signal from each range 
would be digitized by modern 
time-to-digital converters (TDCs), 
e.g. the ATLAS Muon TDC ASIC~\cite{NIMA_453_365} 
whose least significant bit (LSB) resolution is 0.52~ns.
A field-programmable gate array (FPGA) would then choose which range
is most appropriate and send only that one
to be read by data acquisition computers.

In this article, section~2 describes the design of the new QTC ASIC.
Section~3 describes the performance of the ASIC
on the final fabricated chip.
Section~4 presents our conclusions.

 \section{Design of the QTC ASIC}
  \subsection{Overview of the design}

\begin{figure}
 \centering
 \includegraphics[width=8.3cm]{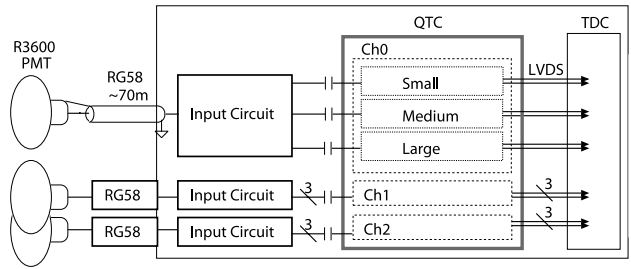}
 \caption{
 Block diagram of the QTC and its surroundings.
 PMT signals, transmitted through coaxial
 cables, are divided among three QTC gain ranges.
 Combined with the input circuits, the QTC provides full analog signal processing for the PMT signals.
 Output signals are generated by LVDS drivers
 and read by TDCs.
 }
 \label{PMT_QTC_TDC}
\end{figure}

Figure~\ref{PMT_QTC_TDC} shows a diagram of the QTC and its surroundings,
and Table~\ref{QTCspec} summarizes its specifications.
  The QTC has three input channels per chip.
  Each channel has three gain ranges: Small, Medium, and Large.
  The gain of each range can be adjusted by external resistor networks. 
  The overall charge dynamic range of the QTC is 0.2 $-$ 2500~pC
  if the gain ratio of three ranges is set to $1:\frac{1}{7}:\frac{1}{49}$.
  Gains and ratios can be optimized to cover a wide dynamic range with
  reasonable resolution.

\begin{figure}
   \centering
   \rotatebox{-90}{
   \includegraphics[height=13cm]{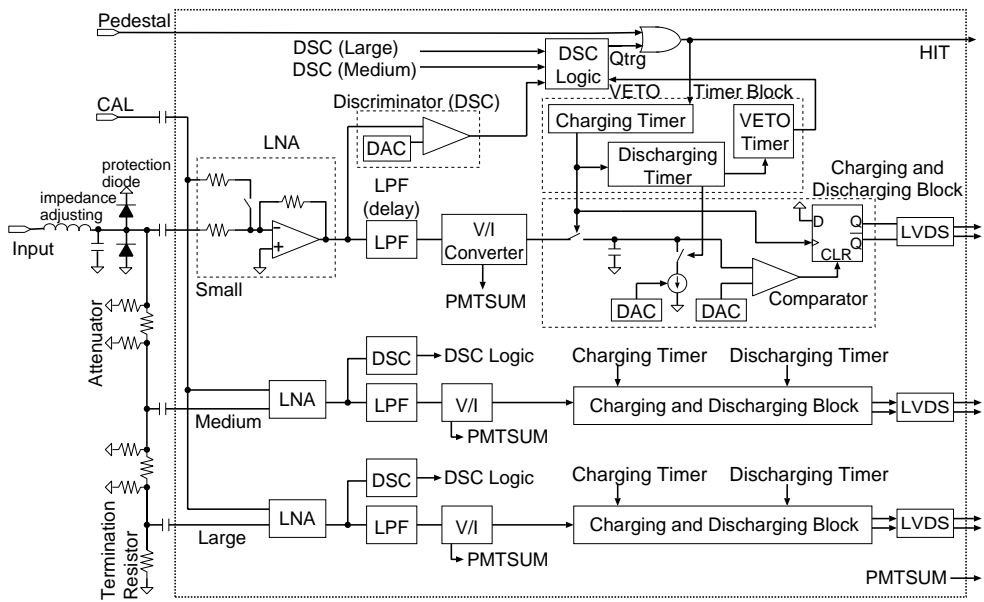}
   }
   \caption{Block diagram of the one QTC channel. 
   Each channel has 3 gain stages: Small, Medium and Large.
   Each stage is composed of several blocks: 
   LNA, LPF, discriminator, V/I converter, and 
   charging/discharging block.
   A single timer block serves all three ranges.
   PMTSUM is the analog sum of the output waveform of
   the V/I converter.
   }
   \label{QTCblock}
  \end{figure}

  Figure~\ref{QTCblock} shows a block diagram of one QTC channel.
  Input signals from PMTs are 
  amplified by a low-noise amplifier (LNA),
  delayed by a low-pass filter (LPF),
  processed by a voltage-to-current (V/I) converter,
  and integrated by a capacitor.
  The sum of the input-signal waveform to the capacitors can be monitored
  through an output signal designated PMTSUM.

  The QTC is of the type of charge sensitive
  and a ramp or Wilkinson's method is used for analog signal conversion.
  Charge integration starts when the amplified input signal crosses
  the discriminator threshold.
  The leading edge of the output signal represents this timing.
  Simultaneously, a trigger flag signal (HIT) is generated.
  After a charging period determined by the timer block,
  the integrated signal starts to be discharged by a constant current
  source.
   The discharging time is proportional to the integrated charge,
   which is known from the width of the output signal.
  Thus, the QTC output signal contains both timing and charge
  information.

  Charge calibration is performed by the calibration signal path (CAL) and
  a forced trigger signal (PEDESTAL).
  Details of CAL are described in Section~\ref{sec:lna}.
  The QTC output signal with null input signal can be measured by asserting 
  the PEDESTAL signal.

 \begin{table}
  \caption{Specification of the QTC}
  \label{QTCspec}
   \begin{tabular}{ll}
    \hline 
    Type of trigger              & self trigger by discriminator\\
    Number of Input Channels    & 3\\
    Processing Speed            & $\sim$ 900~ns/cycle\\
    Charge Integration Gate     & 400~ns\\
    Number of Gains             & 3 (Ratio $1:\frac{1}{7}:\frac{1}{49}$)\\
    Discriminator Threshold      & $-0.3 \sim -14$~mV (small range)\\
    Charge Dynamic Range        & 0.2 $\sim$ 51~pC (small)\\
                                & 1  $\sim$ 357~pC (medium)\\
                                & 5  $\sim$ 2500~pC (large)\\
    Charge Resolution           & $\sim$ 0.2~pC (small)\\
    Integral (Non-)Linearity    & $< \pm$1\% \\
    Timing Resolution           & 0.3~ns (2~pC, -3~mV) \\
                                & $<$ 0.1~ns ($>$ 100~pC)\\
    Power Dissipation           & $<$ 100~mW/ch\\
    Process                     & 0.35~$\mu$m CMOS\\
    Package                     & 100 pin CQFP\\
    \hline
   \end{tabular}
 \end{table}

  \begin{figure}
   \centering
   \includegraphics[width=13cm]{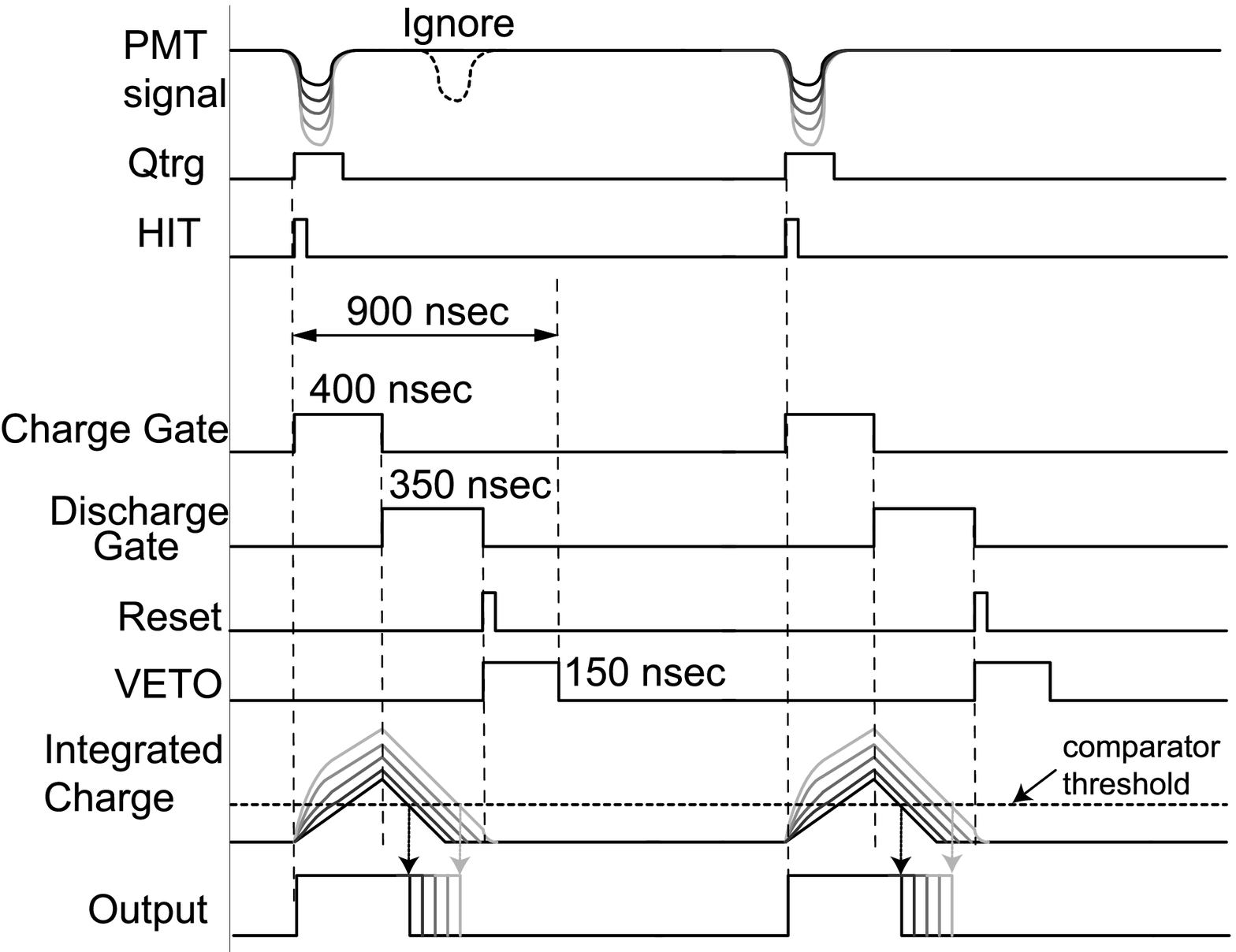}
   \caption{Timing chart for QTC operation.
   Gradation of the PMT signal corresponds to that of the integrated
   charge and the output.
   Widths of the charge and discharge gates are generated by ramps with
   constant discharge current and comparators,
   which can be controlled by register configurations.
   The total processing time for one hit is 900~nsec.
   }
   \label{TimerBlock}
  \end{figure}

  The timer block contains three important timers: 
  a charging timer, a discharging timer, and a VETO timer.
  These timers each generate timing gates by ramps with constant
  discharge current and comparators.
  The time length for these gates are controlled by digital-to-analog
  converters (DACs) for discharge
  current and for comparator threshold levels.

  Figure~\ref{TimerBlock} shows the timing chart for QTC operation.
  The charging timer, triggered by the discriminator output signal,
  operates for $\sim$400~ns.
  Soon after the end of the charge gate, the discharging timer operates
  for $\sim$350~ns.
  During the charge gate, the switch between the  V/I converter and 
  charging capacitor closes and input charge
  accumulates in the capacitor.
  During the discharge gate, the switch between
  the discharging current source and capacitor closes, 
  and arriving input signals are ignored.
  The trailing edge of the QTC output signal represents the time when 
  the integrated signal voltage decreases to the comparator threshold level;
  thus, the QTC output signal widens as the input signal gets larger.
  At the end of the discharge gate, 
  reset and VETO signals are issued.  All QTC circuits
  except the VETO timer are reinitialized and
  the baseline of the LPF output is restored 
  to avoid undesired effects for the next incoming signals.
  Input signals within $\sim$150~ns after the discharge gate are
  ignored.
  In total, processing time for one input signal is $\sim$900~ns.

  Some QTC operations and gate lengths
  can be configured through register settings.
  The dynamic range of the gate length is 150 $-$ 1000~ns.
  The discriminator threshold is controlled by 
  a DAC and by the gain of the discriminator's amplifier.
  The configurable discriminator threshold for the largest gain range
  (Small) is $-0.3$ to $-14$~mV.
  Which discriminator output among the three ranges is
  used to trigger charging
  is also determined by control registers.
  The discriminator threshold can be increased by using a lower gain range  
  for the trigger.
  A channel can be disabled by disabling all three of its ranges.
  The charge dynamic range can be controlled by the gain of
  the V/I converter and the amount of constant discharge current;
  a larger discharge current produces less jitter and has a smaller
  charge dynamic range.
  Discharge current should be optimized to achieve the best resolution 
  over the required dynamic range.

  To minimize power consumption, the QTC is designed to
  operate with a single $+3.3$~V power supply.
  The band gap reference of +1.2~V drives the internal bias circuit
  and built-in 6-bit/8-bit control DACs.

  \subsection{Design of the analog circuit}

  This subsection describes the analog circuits that are
  crucial to QTC performance, 
  to handle accurate impedance matching at the input circuit block,
  wide-band signal amplification in the LNA,
  noise suppression by band limitation at the discriminator block, 
  and signal delay by the LPF for precise charge integration.

   \subsubsection{External input circuit}
   
   As shown in Fig.~\ref{QTCblock},
   there is an external input circuit which 
   contains a termination resistor, gain-adjusting
   attenuators, and DC blocking capacitors 
   outside of the QTC.
   In order to satisfy the requirement for impedance matching
   at precision for $<$0.1\% U-turned reflection signal,
   an external resistor network is used for the input signal termination
   while the QTC input impedance is 10~k$\Omega$.
   This external resistor network is also used to determine the gains of
   the three QTC gain stages.
   In our application for the SK detector,
   input signals are divided by two 17 dB $\pi$ attenuators into the
   three gain stages, for which the ratio of the divided signals is 
   $1:\frac{1}{7}:\frac{1}{49}$.
   Then the divided signals are fed into the QTC via the DC blocking
   capacitors to avoid direct current injection.

   \subsubsection{Low-noise amplifier}\label{sec:lna}

   \begin{figure}
    \centering
    \includegraphics[width=8cm]{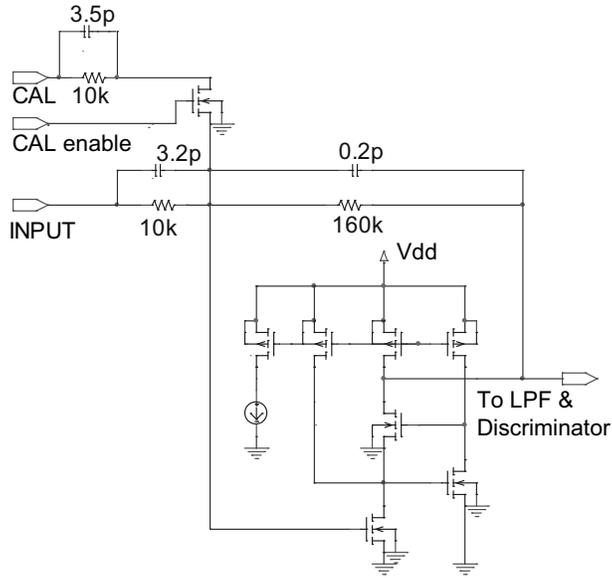}
    \caption{
    Simplified schematic of the LNA.
    The calibration signal path CAL can be switched on and off
    by the MOSFET switch. "CAL enable" is the control register signal
    for the switch.
    }
    \label{LNA_schematics}
   \end{figure}

   Figure~\ref{LNA_schematics} shows the LNA,
   which amplifies the input PMT signals.
   The LNA is a resistor feedback inverting amplifier
   with a cascode stage.
   Resistive rather than capacitive feedback is used
   to handle signals with wide bandwidth.
   The signal path CAL is devoted to charge calibration.
   A common calibration pulse can be supplied to all ranges of all
   channels.  The path can be switched
   on and off by a MOSFET single-pole single-throw (SPST) switch
   controlled by register settings.

\begin{figure}
    \centering
    \includegraphics[width=8cm]{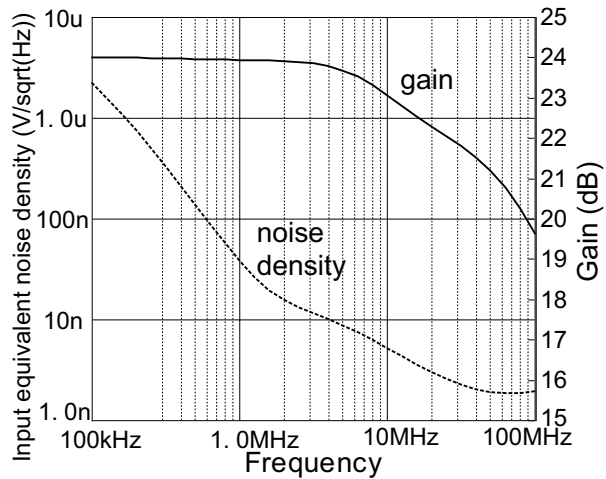}
   \caption{
    Simulated gain and equivalent input noise density
    of the LNA as a function of frequency.
    }
   \label{LNA_freq_response}
   \end{figure}

   Figure~\ref{LNA_freq_response} shows the simulated gain
   and equivalent input noise density of the LNA
   as a function of frequency.
   The low-frequency gain is 24~dB, and the 3-dB bandwidth is 60~MHz.
   In our simulation, PMT signals with a rise time of 10~ns are amplified 
   without shape distortion
   when the quiescent bias current is as low as 2~mA,
   even with a high slew rate.
   The total equivalent input noise amplitude 
   in a bandwidth from DC to 100~MHz
   is estimated to be
   30~$\mu$Vrms, equivalent to 1/100~p.e.s, sufficiently small
   for our applications.

    \begin{figure}
     \centering
     \includegraphics[width=13cm]{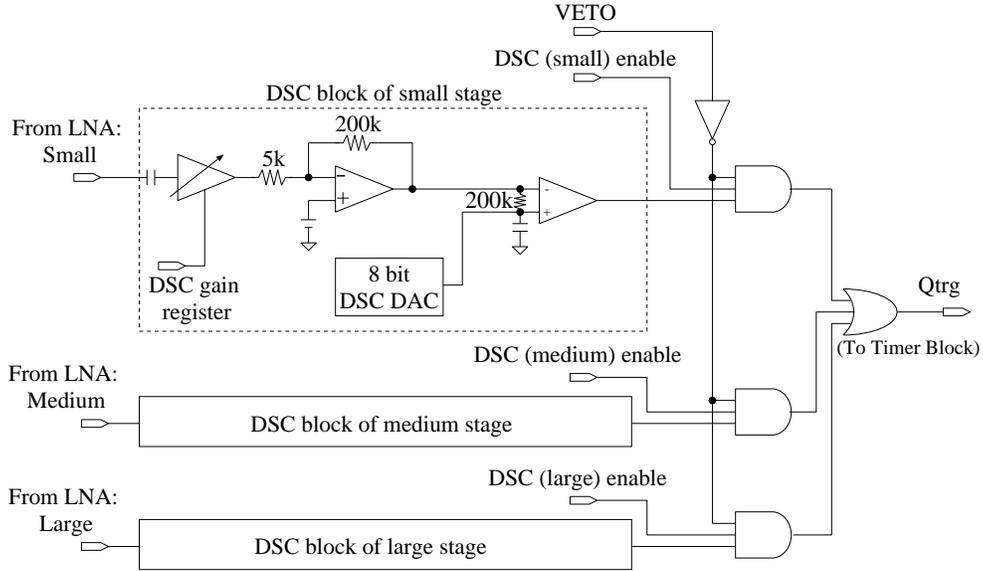}
     \caption{Schematic of the discriminator block.
     Each gain stage has an independent discriminator circuit.
     Which gain stage is used for the trigger source is determined by 
     register settings.
     }
    \label{discriminator_block}
    \end{figure}

   \subsubsection{Discriminator}~\label{sec:dsc}
   
   Figure~\ref{discriminator_block} shows that
   signals amplified at the LNA are fed into the
   discriminator block.
   At the first stage of the discriminator block,
   input signals are amplified again by a variable-gain amplifier and
   a fixed-gain amplifier.
   
    \begin{figure}
     \centering
     \includegraphics[width=8cm]{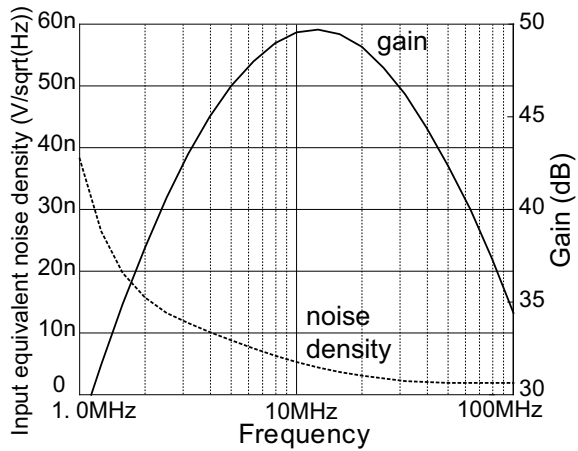}
     \caption{
     Simulated results for discriminator-block amplifier gain
     (including LNA gain) and equivalent input noise density, 
     as a function of frequency. 
     }
     \label{DSC_freq_response}
    \end{figure}

   Figure~\ref{DSC_freq_response} shows the simulated results for gain
   and for equivalent input noise density for the amplifiers in the
   discriminator block.
   To suppress noise density in the lower-frequency region, 
   the bandwidth of the amplifiers is optimized so that the lower and
   upper 3-dB cutoff frequencies are 5 and 30~MHz, respectively.
   The total equivalent input noise amplitude at the comparator is 
   estimated to be 30~$\mu$Vrms, equivalent to 1/100~p.e.s.

   Timing jitter for the discriminator is determined by
   the noise voltage and the slew 
   rate of the input signal.
   The equivalent input noise voltage of 30~$\mu$Vrms is amplified by
   50~dB,
   resulting in an equivalent output noise voltage of 10~mVrms.
   For a noise voltage of 10~mV and a maximum slew rate of 0.1~V/ns,
   we estimate the timing jitter to be 0.1~ns.

   The discriminator threshold can be set by an 8-bit DAC 
   with a step of 10~$\mu$V/LSB.
   The dynamic range of the threshold can be enlarged to a maximum of
   1~V by reducing the
   gain of the variable-gain amplifier and selecting only the lowest gain
   stage (Large)
   for the trigger source.

   \subsubsection{Delay (low-pass filter)}
   
   In a self-triggering scheme, charge integration circuits are
   triggered by the input PMT signal itself.
   The PMT signal to be integrated must be delayed to locate it 
   in the charge gate,
   otherwise not whole charge is integrated.
   In the QTC, we create the necessary delay by using a LPF.

   The LPF must have not only a long delay, 
   but also a shorter tail than the charge gate time of 400~ns.
   A high-order LPF, although ideal for delaying input signals
   without loss of waveform symmetry, is unduly large
   because of its many components,
   especially bulky capacitors.
   Consequently, the QTC uses a second-order
   voltage-controlled voltage source (VCVS) LPF whose
   damping characteristic is comparatively sharp 
   with few components.
   
   \begin{figure}
    \centering
    \includegraphics[width=8cm]{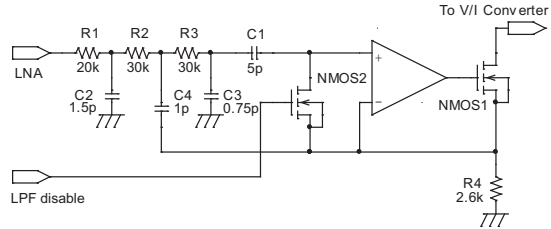}
    \caption{
    Circuit diagram of the VCVS LPF with positive
    feedback provided by the capacitor.
    At the reset period, the "LPF disable" signal closes the NMOS2
    switch, and the baseline of the LPF input is restored.
    }
    \label{VCVS}
   \end{figure}
   
   Figure~\ref{VCVS} shows the circuit diagram of the VCVS LPF.
   The low-pass cutoff frequency is given by
   $f_c = 1/(2 \pi R_3 \sqrt{C_3 C_4}) = 6.1$~MHz.
   The quality factor is given by
   $Q = \sqrt{(C4/C3)}/2 = 0.58$.

   \begin{figure}
    \centering
    \includegraphics[width=8cm]{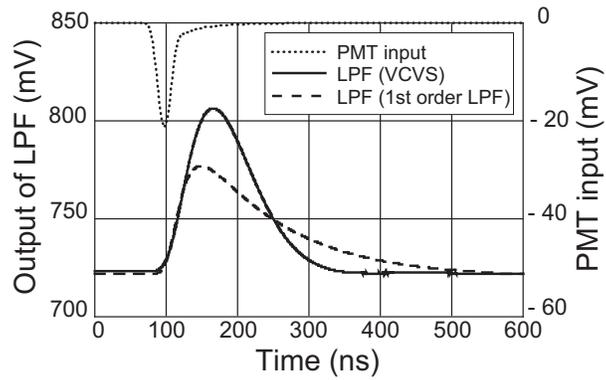}
    \caption{Simulated waveforms for the LPF outputs.
    The delayed waveforms for the VCVS LPF and the first-order passive
    LPF are denoted by solid and dashed lines, respectively.
    }
    \label{LPF_single_pulse}
   \end{figure}

   Figure~\ref{LPF_single_pulse} compares the simulated waveforms for
   the VCVS LPF and first-order passive LPF.
   The VCVS LPF achieves short settling time ($\sim$200~ns, shorter than
   the charge gate time) and
   sufficiently long delay
   ($\sim$20~ns, longer than the sum of $\sim$15~ns rise-time of PMT signals
   and $\sim$2~ns gate delays in the logic needed to open the
   integration gate).
   The output tail of the passive LPF is longer and
   does not settle within the charge gate time.

   The LPF output has equal areas above and below zero volts
   because of the AC coupling capacitor (C1 in Fig.~\ref{VCVS})  on the
   signal line.
   When the pulse repetition frequency is high compared with the lower 
   cutoff frequency 
   of the capacitor, the system suffers signal droop and baseline wander.
   To avoid charge-measurement biases for successive PMT signals,
   we restore the LPF baseline by forcing the signal line to connect
   to ground during 
   the reset period of every processing cycle.

   Since floating capacitors (e.g. C1) are large-area
   metal-insulator-metal capacitors (MIMCaps),
   we should be concerned about capacitive couplings with other signal
   lines.
   The C1 capacitor has an area of 150~$\mu$m $\times$ 45~$\mu$m.
   If a logic line with a width of 0.8~$\mu$m is arranged above the
   capacitor, a parasitic capacitance between them can be as large as
   10~fF.
   Such a large coupling with a logic signal, which rises and falls
   steeply, could cause serious troubles such as a significant error on
   a charge measurement or channel crosstalks.
   The QTC in a developmental stage had a few problems caused by
   unexpected parasitic couplings.
   We studied the effect from parasitic couplings using a Focused Ion
   Beam (FIB), which is one of the microfabrication techniques.
   The mask design of the QTC ASIC was modified to remove the couplings
   revealed by the study.

 \section{Performance}
 We evaluated the final fabricated QTC chip using
a test board on which the QTC ASIC, a TDC~\cite{NIMA_453_365}
and a FPGA are mounted.
This section describes the measured charge dynamic ranges, linearities, 
charge and timing resolutions,
channel isolations, ambient temperature dependences, and rate dependences.

\begin{figure}
  \centering
  \includegraphics[width=8.3cm]{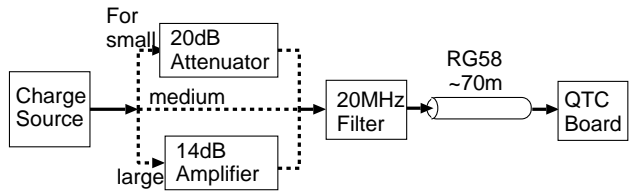}
  \caption{
  Setup for QTC performance evaluation.
  Three signal paths are prepared for the three different gain ranges.
  }
  \label{test_setup}
 \end{figure}

 Figure~\ref{test_setup} shows the performance-evaluation setup.
 For studies on charge linearity and charge/timing resolution,
 we used a calibrated programmable charge source 
 (9.2~fC/LSB, 16-bit dynamic range).
 We prepared three signal paths to achieve a wide range of calibration pulse.
 To create small signals ($<$1~pC, $\sim$1~mV pulse height) with good
 signal-to-noise ratio and linearity, we inserted a high-precision wide-band
 20-dB attenuator into the signal line.
 To create large signals above the dynamic range of the
 charge source,
 we amplified the signal from the charge source by a custom
 charge amplifier with dynamic range up to 3000~pC ($\sim$4~V
 pulse height)
 and nonlinearity within $\pm$1\%.
 At the merging point, 
  higher-frequency components of the charge source signals were filtered
  by a 70-m RG58 coaxial cable and a 20-MHz low-pass filter
  to reproduce the actual rise time of the R3600 PMT signal
  at the SK detector. 
  The charge integration gate was set to 400~ns and the charge dynamic range 
  of the QTC was set to 0.2 $-$ 2500~pC (Table~\ref{QTCspec}).
 The threshold of the discriminator was set to $-0.6$~mV unless
 otherwise noted.

 The noise level of the preamplifier and discriminator was tested
 by counting noise hits in the absence of input signal.
 No noise hits were observed at $-0.3$~mV discriminator threshold,
 equivalent to 0.1~p.e.s and sufficiently below
 the threshold of 1/4~p.e.s for real experiments.

 \begin{figure}
  \centering
  \includegraphics[width=8.3cm]{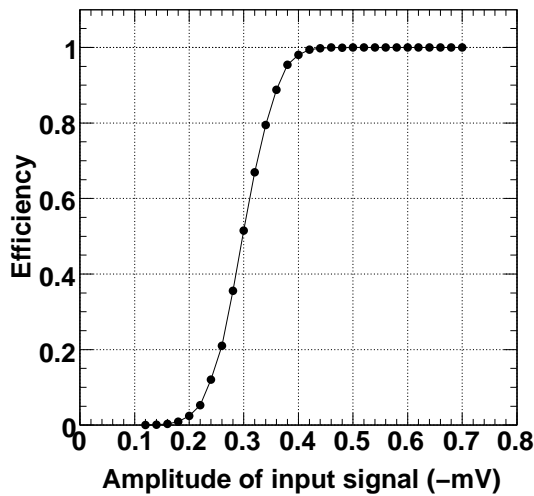}
  \caption{Efficiency curve of the discriminator with $-0.3$ mV (0.1 p.e.s)
  threshold.}
  \label{dsc_curve}
 \end{figure}

 Figure~\ref{dsc_curve} shows a discriminator efficiency curve at the
 threshold.
 The curve's reasonable shape is due
 to the good noise performance of the discriminator.

 \begin{figure}
  \centering
  \includegraphics[width=8.3cm]{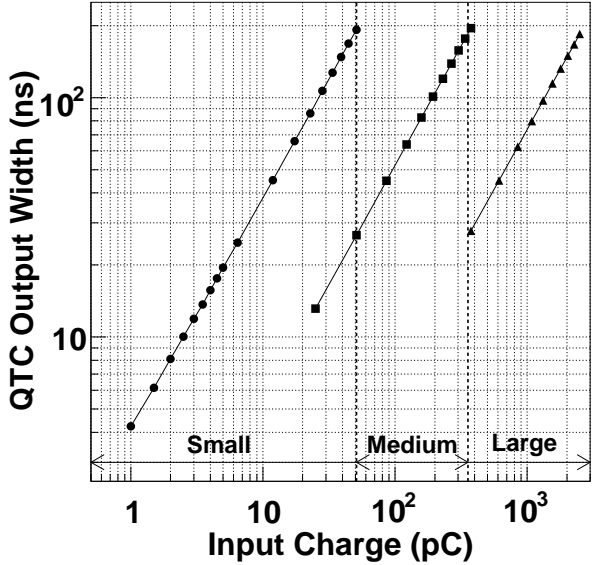}
  \caption{
  Charge responses of the three QTC ranges as a function
  of input charge.  The vertical axis is the width of the output signal
  after subtracting the charging time (pedestal).
  Measurements for the Small, Medium, and Large ranges are denoted by circles,
  squares, and triangles, respectively.
  The lowest gain stage (Large) remains unsaturated up to 2500 pC.
  }
  \label{Q_dynamicrange}
 \end{figure}

 Figure~\ref{Q_dynamicrange} shows outputs
 of the three QTC ranges
 as a function of input charge.
 Each range converts an input charge to an output width 
 with a maximum variance of 200~ns.
 The lowest gain stage (Large)
 remains unsaturated up to $\sim$ 3~V,
 equivalent to 2500~pC.

 \begin{figure}
  \centering
  \includegraphics[width=8.3cm]{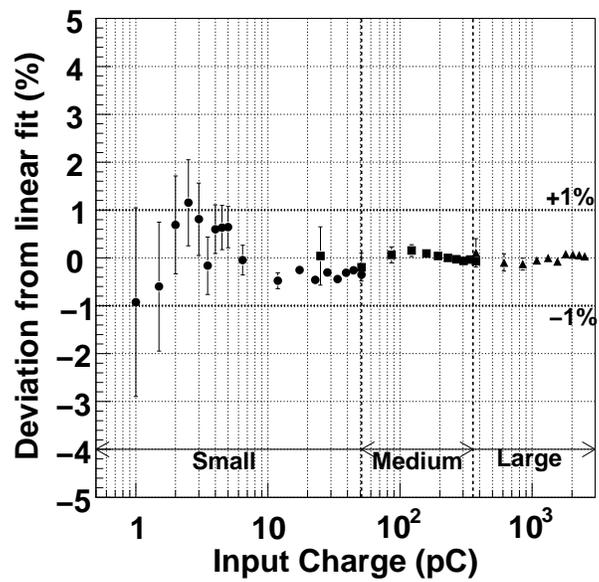}
  \caption{Charge linearity for each QTC gain stage.
  Measurements for the Small, Medium, and Large ranges are denoted by circles,
  squares, and triangles, respectively.
  Nonlinearity is within $\pm$1\%.
  }
  \label{Qlin}
 \end{figure}
   
   Figure~\ref{Qlin} shows the charge nonlinearity for each gain stage,
   where nonlinearity means deviation
   from a linear fitted function.
   The QTC achieves good 
   ($\pm$1\%) charge linearity over a wide dynamic range, 
   and satisfies our requirements.

 \begin{figure}
  \centering
  \includegraphics[width=8.3cm]{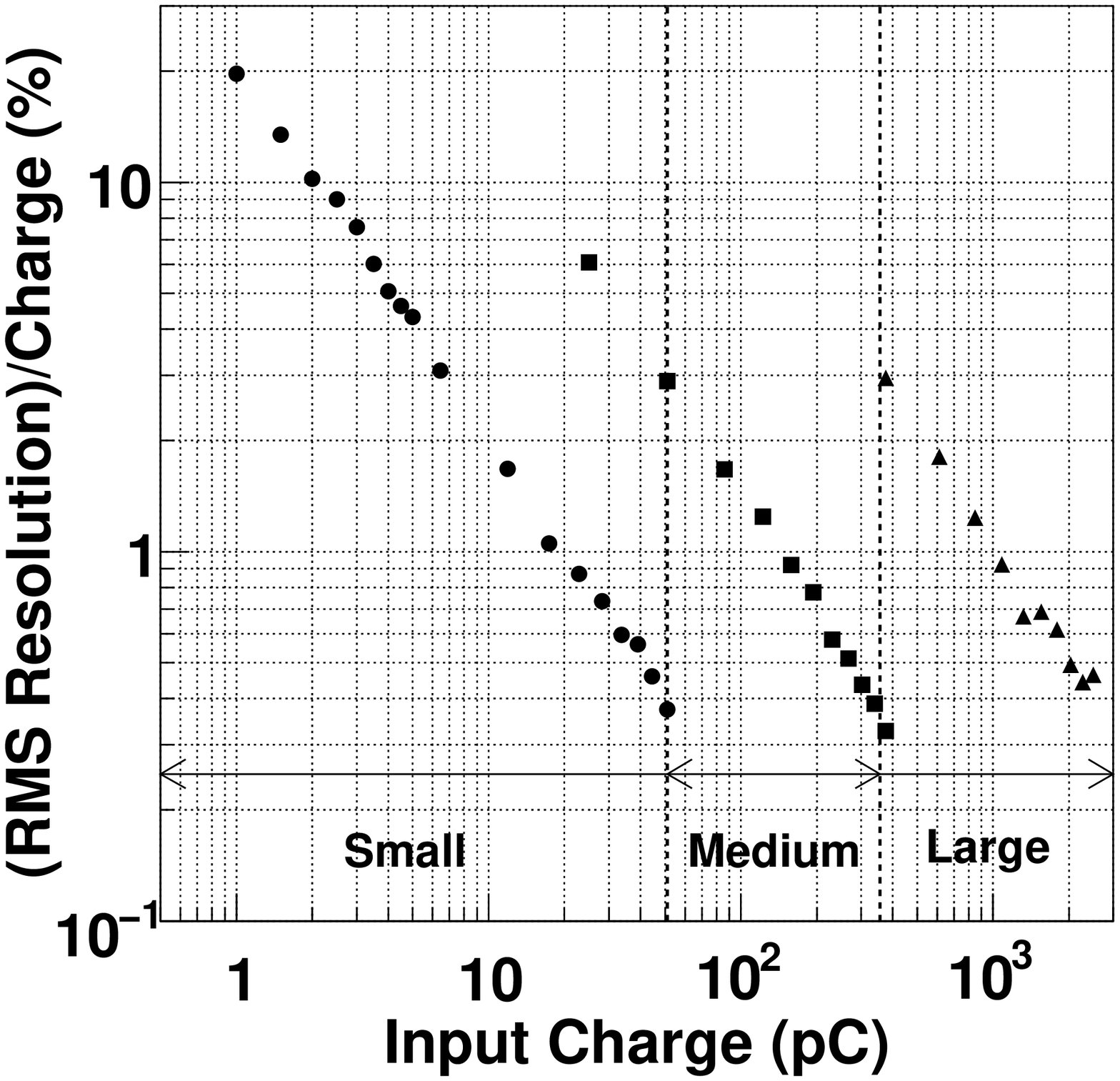}
  \caption{
  Relative charge resolution (percent).
  Measurements for the Small, Medium, and Large ranges are denoted by circles,
  squares, and triangles, respectively.
  The relative charge resolution varies inversely with input
  charge because
  the RMS of the QTC output signal width is almost constant.
  } 
    \label{Qres}
 \end{figure}

 Figure~\ref{Qres} shows that the QTC maintains good resolution over
 a wide dynamic range.  
 The relative charge resolution is $\sim$10\% for a single p.e. level
 ($\sim$2~pC) and decreases to $<$3\% for $\ge$10~pC.
 The RMS absolute values are about 0.2~pC, almost
 independent of input charge.
 These resolutions are sufficiently good compared with 
 the intrinsic charge resolution
 ($\sim$100\%) of the R3600 PMTs.

    
 \begin{figure}
  \centering
  \includegraphics[width=8.3cm]{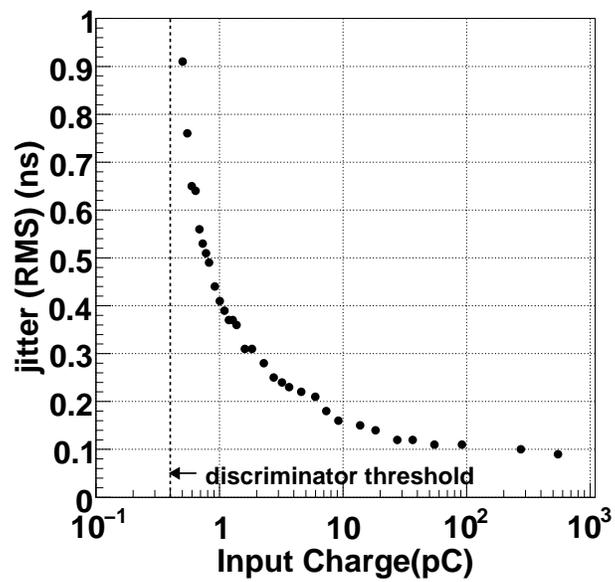}
  \caption{Timing resolution of the QTC output signal.
  The discriminator threshold (dotted line) is $-0.6$ mV,
  equivalent to 0.4 pC.
  The resolution is well below that of the R3600 PMTs, 
  2$\sim$3~ns at $\sim$2~pC.
  }
  \label{tres}
 \end{figure}

 Figure~\ref{tres} shows the timing resolution of the QTC output signal.
 The timing resolution is the RMS of the leading-edge timing,
 measured by a 1-GHz digital oscilloscope.
 The timing resolution for $\ge$100~pC is better than 0.1~ns,
 consistent with that estimated by
 the designed noise voltage and slew rate (Section~\ref{sec:dsc}).
 For smaller signals, resolution increases as the input signal
 amplitude approaches the discriminator threshold.
 When the threshold is $-0.6$~mV, as might be set for
 real experiments,
 the timing resolution is 0.3~ns for the single p.e. ($\simeq2$~pC) level,
 sufficiently better than the intrinsic timing resolution
 (2$\sim$3~ns)
 of the R3600 PMTs.

 Channel isolation is evaluated by counting hits caused by crosstalk
 when an input signal is applied to neighboring channels.
When the discriminator threshold level for a channel is set to $-0.3$~mV 
and $-2$~V input signal is applied to the neighboring channels,
no false hits result from crosstalk.
This is sufficient
because neighboring channels will also have real hits
when a PMT receives a large amount of light.

\begin{figure}
   \centering
   \includegraphics[width=8.3cm]{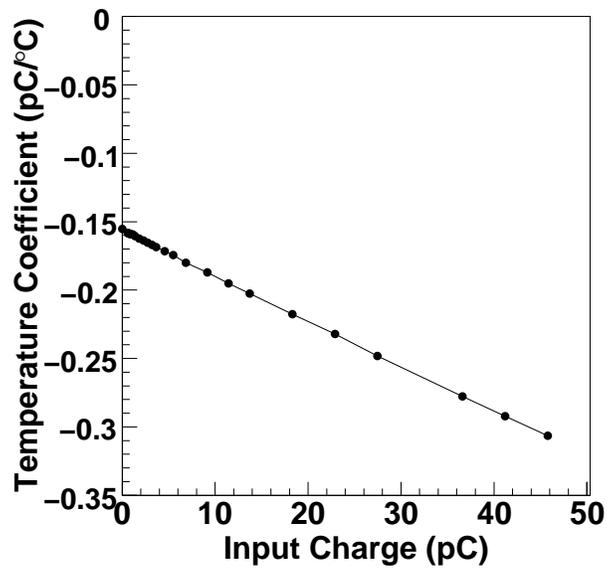}
   \caption{Temperature coefficients for the highest gain stage (Small).
   Coefficients vary linearly with input charge.
   Coefficients for the Medium and Large stages are 7 and 49 times,
   respectively, larger than this.
   }
   \label{tempco}
  \end{figure}

 Figure~\ref{tempco} shows 
 the ambient temperature dependence of QTC charge response,
 measured in a constant temperature bath.
 Output width varies linearly with ambient temperature;
 the coefficients for pedestal data and full-scale charge data
 are equivalent to $-0.15$ and $-0.30$~pC/$^\circ$C, respectively. 
 The temperature coefficient varies linearly with input charge. 
 This linear relationship enables us to correct for possible ambient 
 temperature changes in real experiments by periodically taking pedestal data
 (=0~pC) as well as calibration data at designated input charges.

\begin{figure}
   \centering
   \includegraphics[width=8.3cm]{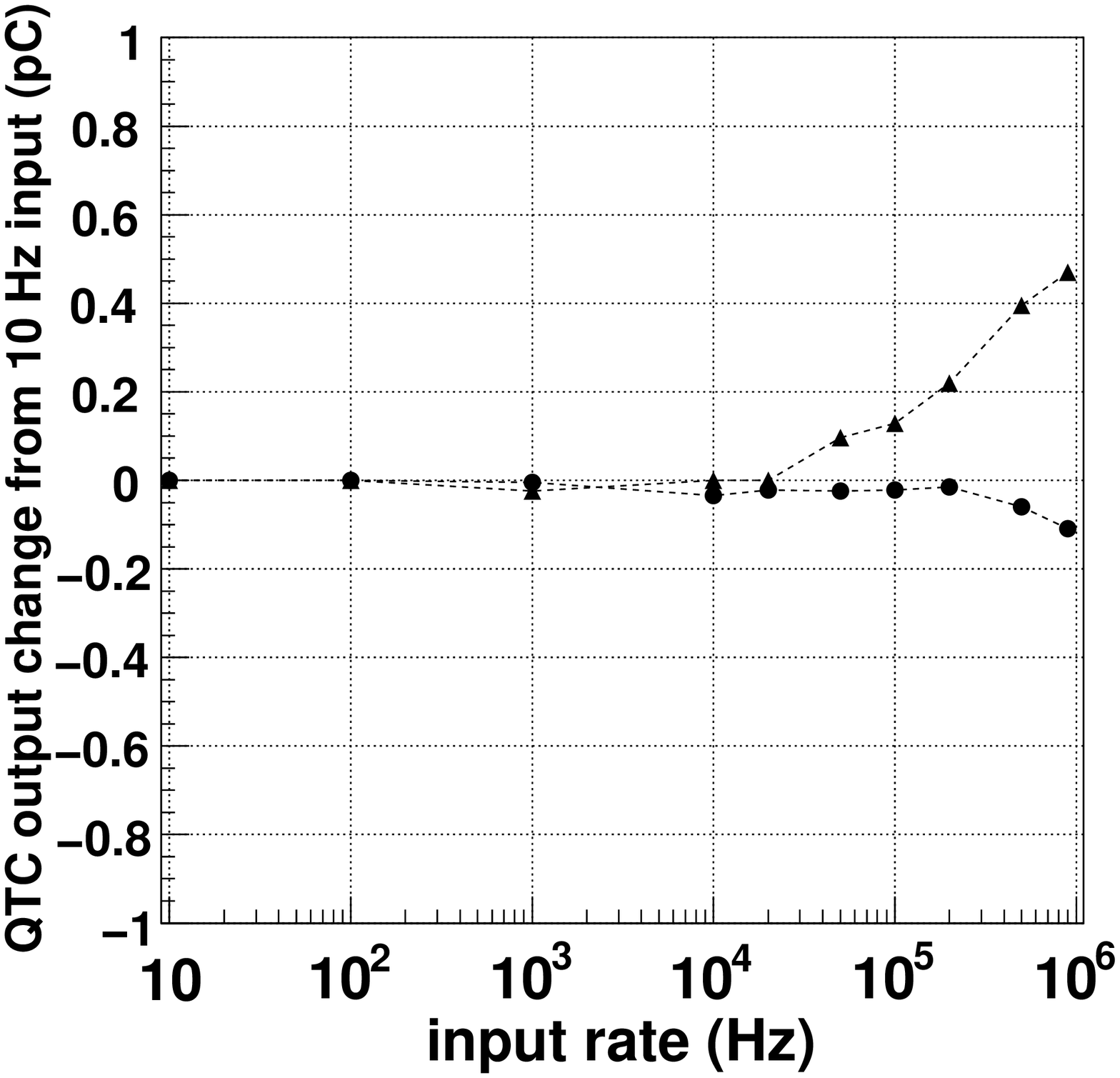}
   \caption{Rate dependence of the QTC output signal
   as a function of input rate.
   Changes of output width are $<$0.5~ns (0.1~pC) and 2~ns (0.5~pC) 
   for input charge of 2~pC (circle) and 51 pC (triangle), respectively.
   }
   \label{fig:rate}
  \end{figure}

  Figure~\ref{fig:rate} shows 
  input rate dependence of the QTC output signal from 10~Hz to 1~MHz.
  For high-speed signal processing, the input rate dependence must be
  sufficiently small.
  For input at the single p.e. (2~pC) level, 
  dependence is $<$0.5~ns (0.1~pC).
  Behavior also depends on the width of the QTC
  output signal in each range.
  The dependence for 51-pC input, which is full scale for the highest
  gain stage (Small), is $<$2~ns, equivalent to 1\% change.
  We thus confirmed that the logic circuits as well as the baseline
  restoration circuit work as designed.

 \section{Conclusion}


A new high-speed QTC has been developed as a mixed-signal ASIC for
PMT readout.
Test results show that the QTC satisfies our requirements.
The charge and timing resolutions for the single p.e. level are 10\% and
0.3~ns, respectively,
both better than the intrinsic resolutions of the PMTs.
Good charge linearity (1\%) is achieved.
The discriminator can be operated with a $-0.3$~mV threshold,
equivalent to 1/10~p.e.s, 
without suffering from intrinsic electrical noise or channel crosstalk.

In September 2008, we successfully replaced the SK data acquisition 
electronics with new front-end electronics on which 
QTCs as reported herein are mounted.
Since then, the SK detector has been stable and has operated well.
Further results on atmospheric/solar neutrinos, high-intensity accelerator
neutrinos, proton decay searches, and more will be reported in the near future.

 \section*{Acknowledgments}
 We would like to thank the Super-Kamiokande collaborators
 for supporting this project and making many helpful discussions.
 A part of this work was carried out at VLSI Design and Education Center
 (VDEC), University of Tokyo.
 This work was partially supported by Research Fellowships of the Japan
 Society
 for the Promotion of Science for Young Scientists.








\end{document}